\documentclass{aims}

\usepackage{txfonts}
\usepackage{amsmath}
\usepackage{tensor}
\usepackage{booktabs}
\usepackage{array}
\usepackage{algorithm}
\usepackage{algpseudocode}
\def\typeofarticle{Research Article}
\def\currentvolume{19}
\def\currentissue{12}
\def\currentyear{}

\def\ppages{12146--12159}
\def\DOI{10.3934/mbe.2022565}
\def\Received{16 May 2022}
\def\Revised{05 August 2022}
\def\Accepted{11 August 2022}
\def\Published{19 August 2022}

\numberwithin{equation}{section}

\begin{document}

\title{Classification of vertices on social networks by multiple approaches}

\author{%
  Hacı İsmail Aslan\affil{1}
  Chang Choi\affil{1}
  and
  Hoon Ko\affil{2,3}\corrauth
}

\shortauthors{the Author(s)}

\address{%
  \addr{\affilnum{1}}{Department of Computer Engineering, Gachon University, Seongnam-daero, Sujeong-gu, Seongnam-si, Gyeonggi-do, Republic of Korea}
  \addr{\affilnum{2}}{Research Institute of Computer, Information, Communication, Chungbuk National University, 8-7, Chungdae-ro 1, Seowon-Gu, Cheongju-si, 28644, Chungcheongbuk-do, Republic of Korea}
  \addr{\affilnum{3}}{GECAD, Instituto Superior de Engenharia do Porto, R. Dr. Antonio Bernardino de Almeida, 431, Porto, 4249-015, Portugal}}

\corraddr{skoh21@chungbuk.ac.kr; Tel: +82-43-261-3140; Fax:\\ +82-43-263-3140.
}

\begin{abstract}
Due to the advent of the expressions of data other than tabular formats, the topological compositions which make samples inter-related came into prominence. Analogically, those networks can be interpreted as social connections, dataflow maps, citation influence graphs, protein bindings, etc. However, in the case of social networks, it is highly crucial to evaluate the labels of discrete communities. The reason underneath for such a study is the non-negligible importance of analyzing graph networks to partition the vertices by using the topological features of network graphs solely. For each of these interaction-based entities, a social graph, a mailing dataset, and two citation sets are selected as the testbench repositories. The research mainly focused on evaluating the significance of three artificial intelligence approaches on four different datasets consisting of vertices and edges. Overall, one of these methods so-called “harmonic functions”, resulted in the best form to classify those constituents of graph-shaped datasets. This paper, it was not only assessed the most valuable method but also determined how graph neural networks work and the need to improve against non-neural network approaches which are faster and computationally cost-effective. Also, this paper showed a limit to be excessed by prospective graph neural network variations by using the topological features of networks trialed.
\end{abstract}


\maketitle

\section{Introduction}

In the current era of social networking, knowledge exchange in any form of information exhibits liaison in between individuals. These social networks may include physical contacts, messages sent, collaborative studies between peers, sentimentally closeness, etc. While the information flow is constructing a network, the format appears as a knowledge map~\cite{I1} which implies a graph structure. Hence, analysis of graph formatted data is essential to obtain more details regarding social networks.

Typically, the data representations tend to be in a tabular or non-Euclidean format to make the data analyzed in a structured way~\cite{I2}. Those data which were expressed in terms of samples and attributions related to every single sample can be imagined as rows and columns intersecting. The former can not imply any relatedness between samples, since tabular-formatted data structure will not provide such an ability. However, linkages in the real world are significant enough to be taken into account while working with samples that have interdependencies in aspects of many topics possibly. Thus, graph networks are the key players here to make a significance by their abilities to show related samples with links in between. In addition to mentioning graph networks, the mathematical foundations of graph networks will be explained in the later sections briefly.

As a variant of the neural network family which has the ability to handle non-Euclidian data, a Graph Neural Network~\cite{I3}, called GNN for the sake of simplicity, is a brand new method in use. Namely said GNNs can take grid-wise graph inputs which show the inter-correlations between samples and evaluate many tasks which will be mentioned later. Having said that, GNNs are getting more popular in aspects of their usage. While its interpretations lead to new approaches, many models are derived depending on GNNs. Its most common fields of achievement are molecular biology~\cite{I4,I5,I6}, network sociology~\cite{I7,I8,I9}, knowledge graphs~\cite{I10,I11}, road traffic~\cite{I12}, natural language processing~\cite{I13, I14}, and even computer vision~\cite{I15,I16}. Adding that, recent development in varied versions of GNNs such as GCN~\cite{I17}, GraphSAGE~\cite{I18}, APPNP~\cite{I19}, SGC~\cite{I20}, GAT~\cite{I21}, DGI~\cite{I22} led this area of neural networks to grow and disseminate in the research field chronologically. Two of these, GCN and GAT, need to be paid attention to in terms of their exemplarity for the other methods since one leads the convolutional approaches whilst the other leads the attention-based mechanisms. Other than variations of GNNs, there should be mentioned other learning methods such as random walks~\cite{I23}, spectral graph theory applications~\cite{I24}, the nearest neighbor approach, and harmonic functions~\cite{I25}. 

The problematics of classifying vertices over a graph by variations of GNNs, whereas non-neural network models are overlooked, influenced this paper. To solve this, GCN and GAT models have been benchmarked against a semi-supervised learning method called harmonic functions. Hereby this article, GCN, GAT, and harmonic functions have been used and the accuracies have been investigated under the same circumstances for each method. Resultingly, this research contributes to understanding the limit to improve graph neural networks and the effectiveness of topological structures while the classification of nodes is aimed.

Our contributions are as follows: (i) We show the topological connections are useful even without node features by using three ML methods for the node classification task while we utilized only topological structures of graph-represented data. (ii) We propose that there is a certain limit of accuracy score to be accessed by prospective modulations of GNNs to be accepted as progressive methods. (iii) We prove the effectiveness of harmonic functions against GCN and GAT when it comes to homogeneous graphs without node and edge features.

\section{Materials and methods}

\subsection{Network datasets and related insights}
\begin{figure}[H]
\begin{center}
\includegraphics[scale=0.8]{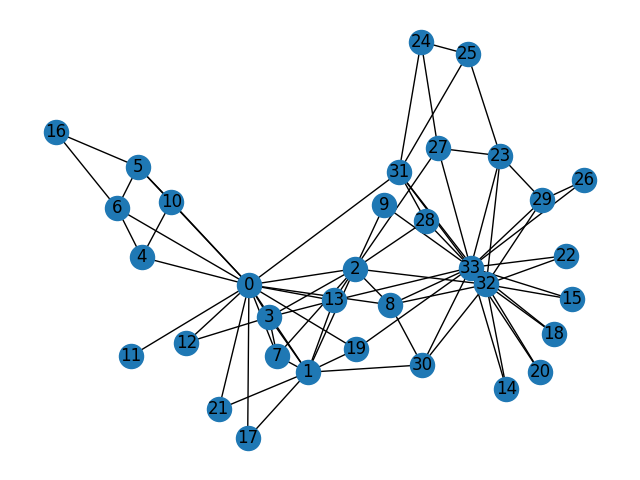}
\caption{Visualization of zachary's karate club dataset.}
\label{Fig2}
\end{center}
\end{figure}

Selection of adequate sets to be subject to processing by particular algorithms is substant for evaluating the throughput of the research. Therefore, the first trials for entire algorithms were performed on a dataset called "Zachary's Karate Club Dataset", created by Zachary et al. in 1977~\cite{M26}. The ground basis for such a selection relies on the simplicity of the dataset mentioned in the previous line. Zachary's Karate Club Dataset, which will be abbreviated as "Karate Dataset" in this article consists of 34 individuals from a local karate club. The dataset shows the members via vertices and the social interactions between club members using edges, basically. Later in time, the group was split into two subsections because of an internal argument between officials and the task here is to predict every student's final decision on whom to get the course from. As depicted in Figure~\ref{Fig2}, class tutors are portrayed by nodes number 0 and 33.

Another data that was extracted from a large European research institution was adopted to be used as a testbench dataset by using a network repository SNAP~\cite{M27}. According to the analysis done by Bharali et al.~\cite{M28}, "Email-Eu-Core network" is found to be a Small-World network that exhibits power law-regime with an assortative mixing pattern on the degree of nodes. In the same research paper, the network was said to be robust against random failures but vulnerable to targeted attacks. Besides, the average degree which means the number of edges compared to the number of nodes was relatively huge considering the same criterion for other datasets introduced hereby in this paper. Thus, the graph-network abbreviated as "Email Dataset" had some differences to be under investigation for node-classification. The challenging structure of the dataset led it to be taken as one of these benchmark datasets in this study.

Lastly in this section, two datasets called "Cora" and "Pubmed" datasets should be mentioned since these are covered in the introduced research. Both sets are consisting of scientific writings and the citation linkages of these articles. As might be expected, papers refer to nodes whereas citations are presented as edges for both citation networks. To fetch these datasets in the form of adjacency matrices, a particular python library, Spektral library~\cite{M29}, has been used.

Those citation networks which are commonly used in the graph theory area were quite important to generalize the outputs of classification results discussed in the subsequent "Results" part. Here, there was a hardship in the case of Pubmed dataset due to its average clustering coefficient. The average clustering coefficient (CC) is a phenomenon that refers to the possibility to draw a triangle by a node and its two distinct neighbors. CC can also be interpreted as the fraction of number of closed triplets by number of all triplets in the network graph.

In an outlined form, one can refer to Table~\ref{table1} as a summary of the datasets utilized during the experiments.

\begin{table}[H]
\begin{center}
\caption{Summary of the datasets}
\begin{tabular}{ccccc} \hline
 &\textbf{Karate}&\textbf{Email}&\textbf{Cora}&\textbf{Pubmed} \\\hline
 \textbf{No. Nodes}& 34& 1005& 2708& 19717\\
 \textbf{No. Edges}& 78& 25571& 5429& 44338 \\
 \textbf{No. Clusters}& 2& 42& 7& 3 \\
 \textbf{No. Training Nodes}& 2& 804& 2166& 5000 \\
 \textbf{Average Clustering Coefficient}& 0.57& 0.4& 0.24& 0.06 \\
 \textbf{Average Degree}& 4.59& 25.44& 3.9& 4.5 \\\hline
\end{tabular}
	\label{table1}
\end{center}
\end{table}

\subsection{Preprocessing node and edge embeddings}
Most generally, when one's working on graph data without node and edge attributions, the preprocessing schedule should be short. For instance, in the case of the karate dataset, there was no action needed to process the inputs since those were already in an abstract form. Similarly, the email dataset didn't demand anything besides turning the data list to tuples to be handled.

However, for the Cora dataset, there were some necessary transformations to be considered. The main reason why it matters to rebuild the Cora dataset is that the harmonic function takes the node inputs as integers and those integers should be in an adjacent format. Additionally, GAT and GCN take the label arguments as integers as well. Contrary to the requirements, the Cora dataset accommodates vertices in a non-consecutive order. Having said that, particular numbers of nodes are not varying in between the range of numbers of samples which means node numbers should be arranged in order to have those nodes adjacently increasing from zero to the number of total vertices in the set. As $V^{}_i{} \rightarrow V^{'}_i{} \ \ \textrm{and}
\ \ V^{}_j{} \rightarrow V^{'}_j{} \ \ \textrm{(where $V^{}_i{} \ \ \textrm{and} \ \  V^{}_j{}$\ \ refers to the vertices having an interconnection )}$ the edge connection should remain the same as $Edge(V^{}_i{},V^{}_j{}) \rightarrow Edge(V^{'}_i{},V^{'}_j{})$ while $Edge$ showing the link between two vertices.

\begin{table}[H]
\begin{center}
\caption{Original label names and related integers while preprocessing.}
\begin{tabular}{cc} \hline
 \textbf{Original Label}&\textbf{Encoded Target} \\\hline
 Rule Learning& 0\\
 Neural Networks& 1\\
 Theory& 2\\
 Case Based& 3\\
 Probabilistic Methods& 4\\
 Genetic Algorithms& 5\\
 Reinforcement Learning& 6\\\hline
\end{tabular}
	\label{table2}
\end{center}
\end{table}

The aforementioned situation of labels of the cora dataset has been solved by mapping the clique texts to specific integers. One can refer to Table~\ref{table2} for the encoding details per each label name.

\subsection{Graph Convolutional Networks and Graph Attention Networks}
To fulfill the aim of this study, two networks were applied to the datasets and described in the following subsections. It should be kept in mind that many other models could be used to achieve classification tasks. However, in this paper, when comparing harmonic functions with GCN and GAT, it is aimed to set a higher limit in terms of accuracy than the lower limit that harmonic functions will define for future neural network models to achieve.
\subsubsection{Graph Convolutional Networks}
Proposed by Kipf et. al.(2017), Graph Convolutional Networks~\cite{M81} can be seen as the enhancements on neural networks that operate on graphs which had been previously introduced by Gori et al. (2005)~\cite{M82}. For convenience, it would be nifty to show the mathematical foundations of GCNs so that the surrounding dialectics can be understood well.

Prescribed as inputs \textit{\textbf{A}}, the adjacency matrix in the shape of $\textit{\textbf{N}} \times \textit{\textbf{N}}$, and \textit{\textbf{H}}, the feature matrix in the shape of $\textit{\textbf{N}} \times \textit{\textbf{F}}$, where each vertex has 2-D feature vector in the shape of $\textit{\textbf{1}} \times \textit{\textbf{F}}$, the basic propogation rule for a GCN is given by Equation~\ref{equation1}.

\begin{equation}
\textit{H'}=\boldsymbol{\sigma} \textit{(AHW)}
\label{equation1}
\end{equation}

Noting that W is a the weight matrix in the shape of $\textit{\textbf{F}} \times \textit{\textbf{F'}}$, \textit{\textbf{H'}} stands for the newly generated node features, and $\boldsymbol{\sigma(\cdot)}$ implements the non-linearity function, the node \textit{i} was focused to get the following equation to simply sump up the transformed features of all connecting nodes. Equation~\ref{equation2} depicts the correlation in between a node's features $h_i$ and the features of the connected nodes $h_j$ to the particular node \textit{i}. Please note that $\textit{\textbf{V(i)}}$ is the group of node \textit{i}'s neighboring nodes.

\begin{equation}
h_i = \sigma \left( \sum_{j\in \textit{V(i)}} W^{T}h_j \right)
\label{equation2}
\end{equation}

Conflicting with the expression made, the previous propagation rule which is depending on sum-pool doesn't perform solid. The reason which leads to such an outcome is how the propagation rule introduced in Equation~\ref{equation2} just sums up feature vectors. This may lead the repeated applications to increase the scale of the features. To overcome that obstacle, another update rule that can be simulated by Equation~\ref{equation3} satisfies the normalization of the adjacency matrix by multiplying it by the inverse of the diagonal degree matrix \textit{\textbf{D}}. Consequently, a newly updated node feature vector is shown by Equation~\ref{equation4}. 

\begin{equation}
\textit{H'}=\boldsymbol{\sigma} \textit{($D^{-1}$AHW)}
\label{equation3}
\end{equation}
\begin{equation}
h{^{'}_{i}} = \sigma \left( \sum_{j\in \textit{V(i)}} \frac{1}{|V(i)|} W^{T}h_j \right)
\label{equation4}
\end{equation}

The previous propagation rules were presented in order to explain how GCN works. Now, the rule which is derived from the previous equations, so-called symmetric normalization should be visited as Equation~\ref{equation5} shows, to see how it multiplies the adjacency matrix by the square root of the inverse of the diagonal degree matrix. Resultingly, the node features are up to be changed according to Equation~\ref{equation6}

\begin{equation}
\textit{H'}=\boldsymbol{\sigma} \textit{($D^{-1/2}$A$D^{-1/2}$HW)}
\label{equation5}
\end{equation}
\begin{equation}
h{^{'}_{i}} = \sigma \left( \sum_{j\in \textit{V(i)}} \frac{1}{\sqrt{|V(i)|}\sqrt{|V(j)|}} W^{T}h_j \right)
\label{equation6}
\end{equation}

The implementation of GCN in the case of this research has exploited the theoretical foundations expressed in this section. Even so, this study aimed to work with topological bonds of the graph-structured data without fetching any information related to nodes. Hence, features for each node were set to be zero initially. 

\subsubsection{Graph Attention Networks}

Graph Attention Networks (GATs) were introduced in Veličković et al. (2018) as a novel approach that benefits the GCN's background but adds particular "attention" mechanisms. The mathematical pinnings of GAT can be viewed on its inventor's paper~\cite{I21}, though it is still needed to outline the unique annexes of GAT in the context of this paper.

Unlike GCN, the coefficients in GAT are not constant due to the relaxation of the coefficients being dependent on the current input. The idea explained is the motivation of attention mechanisms for such a graph neural network. Having said that we have non-constant coefficients, the attention coefficient $\alpha_{ij}$ while the \textit{j} is the sender and \textit{i} is the receiver nodes were computed as Equation~\ref{equation7} shows.

\begin{equation}
\alpha_{ij} = \frac{\textrm{exp}(\textrm{LeakyReLU} (a[W^{T}h_i||W^{T}h_j]))}{\sum_{k\in \textit{V(i)}} \textrm{exp}(\textrm{LeakyReLU}(a[W^{T}h_i||W^{T}h_k]))}
\label{equation7}
\end{equation}

Theoretically, a one-layered MLP, which is symbolized by \textbf{a}, has been applied on concatenated messages $W^{T}h_i$ and $W^{T}h_i$ by the activation function which was propositioned as the LeakyReLU function. Over and above that, GAT exerts multi-head attention which means each GAT layer has a fixed number of independent duplicates. Those outputs obtained through each duplicated layer of GAT, then, were concatenated to produce a finalized feature vector. As Equation~\ref{equation8} expresses, the normalized attention coefficients are used to determine the features corresponding to them, to assign the final output features for every node. 
\begin{equation}
h{^{'}_{i}} = \sigma \left( \sum_{j\in \textit{V(i)}} \alpha_{ij} W^{T}h_j \right)
\label{equation8}
\end{equation}
\subsection{Harmonic Functions}

As stated by Zhe et al.~\cite{M30}, semi-supervised learning using Gaussian fields and harmonic functions is applicable for networks whose topology is known. With the help of the NetworkX library's useful application programming interface~\cite{M31}, the classification method has been applied to the datasets accordingly. One can refer to the Equation~\ref{equation9} which explains what harmonic functions are where a function $\textit{h} : V \rightarrow \mathbb{R}$ is called harmonic function so that the graph \textit{G = (V, E)} is harmonic. For convenience, \textit{$d_i$} refers to the degree of the vertex. Sufficient information can be revisited from He et al.~\cite{M32} to understand the basis of classification by harmonic functions.

\begin{equation}
h(V{^{}_{i}}) = \frac{1}{d{^{}_{i}}} \sum_{(V{^{}_{i}},V{^{}_{j}})\in V} h(V_j)
\label{equation9}
\end{equation}

\section{Proposed Method}
As its subcomponents are defined in the previous section, basically, the proposed method depends on accomplishing node classification tasks through both GCN, GAT, and harmonic functions. Since one of the main motivations of this article is to show the impact of topological bindings on classification, the very first step of our proposed method to conduct this research is to sift network data from node and edge features to access only topologic attributions. Consequently, there will be a drastic decrease which may lead to more efficient use of memory in terms of the amount of data.

\begin{figure}[H]
\begin{center}
\includegraphics[scale=0.7]{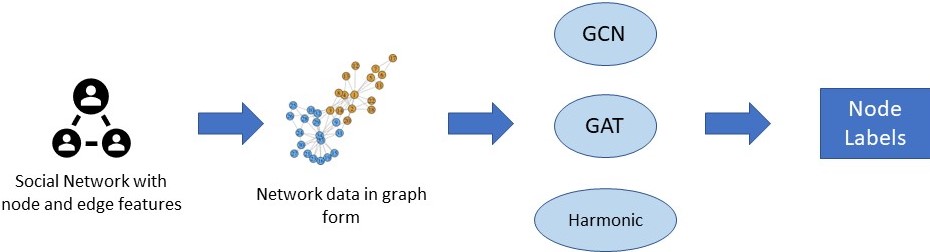}
\caption{Workflow of the proposed method.}
\label{Figx}
\end{center}
\end{figure}

After the extraction of topologic bindings, the aforementioned classification methods were subject to be used separately. Predicted node labels were noted for every classification method to evaluate the importance of the topological structure of graphs for vertex classification and the best model overall for such an application. Figure~\ref{Figx} outlines the projected research method which leads this study. After following those steps depicted in Figure~\ref{Figx}, the rest is to interpret the results accordingly. To have a better perception, Algorithm~\ref{alg:method} has been extracted to outline the general workflow of the proposed study.

\begin{algorithm}
\caption{Concise algorithm of the proposed method}\label{alg:method}
\begin{algorithmic}
\Require $G(V, E, F)$
\Ensure $acc1, acc2, acc3$
\Function{preprocess}{$G(V, E, F)$}:
\State $i \gets 0$
\While{$V_i \in V$}
    \State delete $F_i$
    \State $i \gets i+1$
\EndWhile
\State \Return G(V, E)

\EndFunction

\Function{classify}{$G(V, E)$}:
    \State $acc1 \gets GCN(G(V, E))$
    \State $acc2 \gets GAT(G(V, E))$
    \State $acc3 \gets Harmonic(G(V, E))$
\State \Return acc1, acc2, acc3

\EndFunction
\State $acc1, acc2, acc3 \gets classify(preprocess(G(V, E, F)))$
\State $best\_model \gets max(acc1, acc2, acc3)$

\end{algorithmic}
\end{algorithm}

\section{Results}
\subsection{Environmental Setup}

While the learning rate is 0.01, the optimizer is Adam, loss function depends on a negative log-likelihood approach, out of 10 training sessions with 500 epochs in every case, GCN and GAT methods were trialed whereas harmonic function was only trained within 100 iterations because, after a particular number of iterations, the harmonic function doesn't perform any better or worse compared to previous iterations. At the end of those 10 experiments for each dataset and method, the standard deviations were noted since the training data were sampled randomly. For a better generalization, it has been thought that random sampling would give a better idea. Because serving as a validation dataset for the algorithms, the only case in which the training data was not set randomly was the karate dataset case, which took only 2 node assignments as the training data.

In terms of computing architecture, a cloud computing service so-called Google Colab has been utilized in this research because of its simplicity. Python 3 Google Compute Engine backend (TPU) served as the processor, while 35 GB of memory has been subject to be used.

\subsection{Research results}
\begin{figure}[H]
\begin{center}
\includegraphics[scale=0.27]{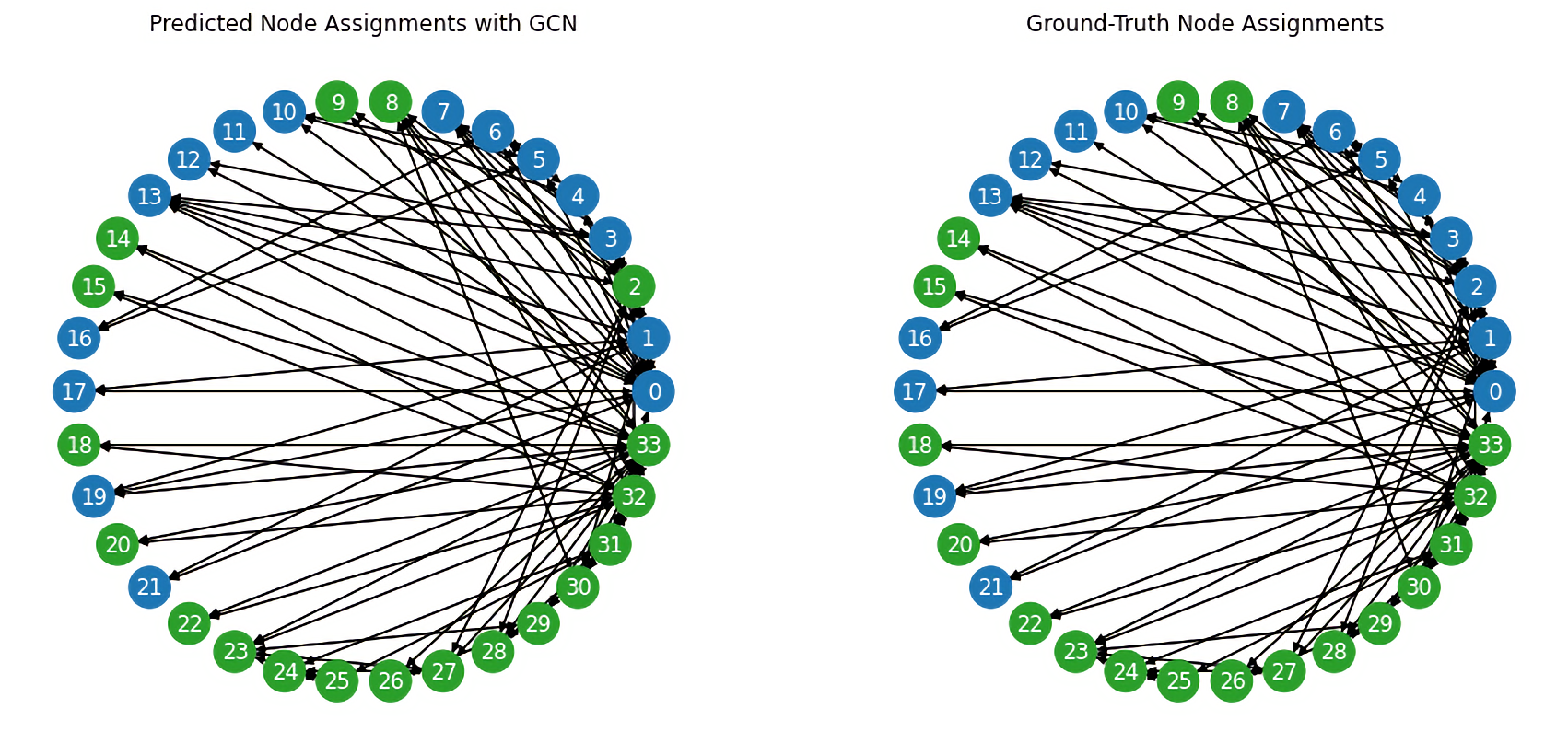}
\caption{Classification results of intentionally mistuned GCN for karate dataset.}
\label{FigZachResult}
\end{center}
\end{figure}

Ensuing the construction of methods and datasets, the very explainable sequel occurred for the karate dataset. Hence, the social interaction in between the individuals herewith helped the experiment to distinguish those two distinct cliques. Figure~\ref{FigZachResult} illustrates the classified labels per node by GCN and nodes' corresponding ground-truth assignments.
 
In the case of karate dataset, it was observed that the classification accuracy could hit the \%100 quite easily without the need of complex parameter tuning, since the dataset is quite small and partaking in a particular clique is linearly dependent to the social interactions of the club members. Therefore, to highlight the difference in terms of accuracy between classified labels and ground-truth assignments on Figure~\ref{FigZachResult}, the GCN was trained during 20 epochs intentionally. In the case of a proper training, the full accuracy was observed.

\begin{table}[!htbp]
\centering
\caption{Comparison of methods used in this research.}
\begin{tabular}{*4c}
\toprule
Name of the dataset &  \multicolumn{3}{c}{Accuracy for full dataset (ACC)} \\
\midrule
{}   & GCN & GAT & Harmonic  \\
\cline{2-4}
Karate  & 100\%& 100\%& 100\% \\
Email   & 94.42±0.52\%& 93.59±0.47\%& 94.14±0.68\% \\
Cora   & 96.29±0.3\%& 96.61±0.29\%& 97.27±0.23\% \\
Pubmed   & 85.29±0.26\%& 85.55±0.24\%& 96.47±0.07\% \\
\midrule
{} & \multicolumn{3}{c}{Accuracy for unlabelled data (test accuracy)}\\

\midrule
{}   & GCN & GAT & Harmonic  \\
\cline{2-4}
Karate & 100\%& 100\%& 100\%\\
Email & \textbf{72.1±0.52\%}& 67.95±0.47\%& 70.7±0.68\%\\
Cora & 81.45±0.3\% & 83.05±0.29\% & \textbf{86.35±0.23\%}\\
Pubmed & 80.39±0.26\%& 80.73±0.24\%& \textbf{82.35±0.07\%} \\

\bottomrule
\end{tabular}
\label{tablex}
\end{table}

In addition to computing the standard deviation, the mean accuracy levels have been measured by simply taking the average of results in 10 experiments. The train to test ratio was 0.8 for every case but Pubmed-GCN and Pubmed-GAT duos. The reason is that the out of memory error occurs since the number of processed nodes is too much in the training for the Pubmed dataset. To avoid such an error, the Pubmed dataset has been trained with 5000 vertices which means almost 25\% of the entire network.

Interpreting the tables in this section is only possible once one knows the train-test split method that has been used in the current paper. The foremost fact about measuring the test accuracy here depends on the training backbones of semi-supervised learning by maximizing the negative log-likelihood of the known node assignments. Here, the mentioned "known node assignments" can be elucidated as train data, and resultingly, an analogy between unlabelled data and test data can be constructed. Therefore, once the model is fed with a certain number of samples whose ground-truth labels are known, the whole graph is taken as an entire set that includes both test and train data, and the accuracy for the entire set can be recorded by Table~\ref{tablex} and abbreviated as ACC. Yet, the observed accuracy contains both train and test accuracy. To evaluate the test accuracy for unlabelled data, a simple fraction of the train to test is used by assuming the training accuracy 100\% as explained in Equation~\ref{equation10}.

\begin{equation}
\textrm{Test Accuracy} = \frac{\textrm{ACC} - \textrm{(Training Samples Ratio\%)}}{{1-\textrm{(Training Samples Ratio\%)}}}
\label{equation10}
\end{equation}

According to the latter formula, accuracies of prediction for unlabelled nodes have been shown by Table~\ref{tablex}.

\section{Discussion}

Dominantly appearing, the harmonic function outputs slightly better than the others for node classification tasks on Cora and Pubmed datasets in terms of mean accuracy levels. Having said that, the standard deviation occurred smaller than it occurred for the other datasets which means harmonic functions are more robust against different training samples. When the number of samples is more, the standard deviation goes lower as expected theoretically. The same or the opposite can not be said for the mean accuracy levels since they depend on internal knowledge in each graph. The amount of samples is only one fact but not the most crucial. When it comes to graph-structured data, the heterogeneity~\cite{D1} issue emerges as the key.

Known as message passing algorithms, both GCN and GAT depend on the same theory but differ on update and aggregation rules. Contingent with these methods, varied versions of usage-specific graph neural networks have been deployed in recent years as stated in the Introduction section. Hence, GCN and GAT served as seed points for many others to initiate fresh methods to overcome graph-related tasks in artificial intelligence.

Though graph neural networks show impressive performance on graph-related tasks in the machine learning field, they still need to exceed certain limits. For sure, most of the cutting-edge techniques have already gone beyond limits previously achieved by GCNs and GATs. However, most of the novel approaches in GNNs still depend on GCN and GAT theory. Intuitively, there must be a comparison level for these new models to be accepted as an improvement. In this research, the limit has been proposed as classification by harmonic functions as introduced by Zhu et al. For the popular datasets, such as Cora and Pubmed datasets, this theory seems to work fine. Unlike the motivation of this theory, GCN outperformed harmonic functions for the Email dataset, which showed us this limit may not be applicable for every dataset.

Due to heterogeneity in graphs, generalization of the limit to be excessed by new approaches as the output of harmonic functions is not possible. However, it can still be set as a limit for particular datasets as has been shown in this paper. Having said that, one should note that only the topological attributes have been used which makes the case out of the natural language processing subject. For further studies, node features will be included for a larger set of test cases. Moreover, edges with attributes may also be included to enhance the topological information yielded by a graph structure.

Whereas harmonic functions slightly outperform GCN and GAT, it still has lack handling feature vectors related to entries. Thus, classification with harmonic functions is only possible with topological correlations of nodes in the graph. This disability of not being able to work with feature matrix can be offered as a soft spot for harmonic functions which need to be fixed by new modifications. Recalling the fact that both GCN and GAT can work with feature matrices as introduced by Equation~\ref{equation1}, merging the node update rule of harmonic functions and the feature update rule of GCN or GAT would devise a new and even better method for node classification tasks. Even so, this idea needs confirmation by further research but still worths to note here in this section.

\section{Conclusion}

Hereby this paper, while using three methods on four sample spaces, the topological bindings of individuals in each sample space have been processed to classify nodes into related communities/labels. The contributions, as stated in the introduction section, was achieved to some extent. The main motivation was reaching a certain limit of accuracy score to be accessed by prospective modulations of GNNs and it is fulfilled for two of sample spaces which are highly used as testbench datasets. Moreover, after sifting data to obtain only topological structure, we could show that inter-correlation data is still useful on its own as a feature for graph neural networks and harmonic functions. Lastly, as our main motivation depicts, harmonic functions appeared more effective against GCN and GAT while working with graphs without node and edge attributions.

\section{Future Work}

The process of preparing the current research article led to many ideas regarding Graph Neural Networks. As it was dived into details, GNNs' handiness to obtain node embeddings like a feature extractor or as an encoder influenced us to investigate if GNN models can be merged with other ML techniques for industrial applications. Especially, the banking transaction datasets that consist of licit and illicit transactions attracted our attention, since these sorts of data refer to time-series data and such applications can be surveyed in the security area. Hence, the authors of this research will look for potential advancements in GNNs by using fusing GNNs with other ML or graph theory techniques to solve industry-related issues.

\section*{Acknowledgments}

This work was supported by the National Research Foundation of Korea (NRF) grant funded by the Korea government (MSIT) (2021R1A2B5B02087169), and Basic Science Research Program through the National Research Foundation of Korea (NRF) funded by the Ministry of Education (No. 2021R1I1A3040361).

\section*{Conflict of interest}

The authors declare there is no conflict of interest.

\end{document}